\documentclass[showpacs,physrev,aps,twocolumn,superscriptaddress,nofootinbib]{revtex4-2}
\pdfoutput=1

\newcommand{\beq}{\begin{equation}}
\newcommand{\eeq}{\end{equation}}

\usepackage{amsmath,amsfonts,amssymb,bm,graphicx,hyperref}
\usepackage{color}
\usepackage{array}

\newcommand{\rhobar}{\bar\rho}

\newcommand{\lambdaC}{\lambda_{\rm c}}
\begin{document}

\title{Mechanical analysis of a dynamical phase transition for particles in a channel}

\author{Jakub Dolezal}
\affiliation{DAMTP, University of Cambridge, Centre for Mathematical Sciences, Wilberforce Road, Cambridge CB3 0WA, United Kingdom}
\author{Robert L. Jack}
\affiliation{DAMTP, University of Cambridge, Centre for Mathematical Sciences, Wilberforce Road, Cambridge CB3 0WA, United Kingdom}
\affiliation{Yusuf Hamied Department of Chemistry, University of Cambridge, Lensfield Road, Cambridge CB2 1EW, United Kingdom}

\newcommand{\citeMFT}{\cite{Bertini2015}}
\newcommand{\citeAppert}{\cite{Appert2008}}  
\newcommand{\citeJackHU}{\cite{Jack2015}}
\newcommand{\citeBias}{\cite{Lecomte2007,Garrahan2009,Jack2020ergo}}  %
\newcommand{\citeJakubMany}{\cite{Dolezal2019}}
\newcommand{\citeIK}{\cite{irving1950}}
\newcommand{\citeLecomteInact}{\cite{Lecomte2012}}  
\newcommand{\citeParola}{\cite{parola2019}}
\newcommand{\citeSchofield}{\cite{Schofield1982}}
\newcommand{\citeSmith}{\cite{Smith2017}}

\newcommand{\eqEOM}{\ref{equ:eom}}
\newcommand{\eqDefC}{\ref{equ:C-def}}
\newcommand{\eqDefBias}{\ref{equ:bias}}
\newcommand{\eqBal}{\ref{equ:balance}}

\newcommand{\figRho}{\ref{fig:rho}}
\newcommand{\figStress}{\ref{fig:stress}}
\newcommand{\figVBE}{\ref{fig:doob}}

\begin{abstract}
We analyse biased ensembles of trajectories for a two-dimensional system of particles, evolving by Langevin dynamics in a channel geometry.
This bias controls the degree of particle clustering.  On biasing to large clustering, we observe a dynamical phase transition where the particles break symmetry and accumulate at one of the walls.  We analyse the mechanical properties of this symmetry-broken state using the Irving-Kirkwood stress tensor.  The biased ensemble is characterised by body forces which originate in random thermal noises, but have finite averages in the presence of the bias.  We discuss the connection of these forces to Doob's transform and optimal control theory.
\end{abstract}

\maketitle
 
 \section{Introduction}
 
Large deviation theory~\cite{denH-book,dupuis-book,Touchette2009} is now an established tool for analysing dynamical fluctuations and rare events in physical systems.  Fluctuation theorems and thermodynamic uncertainty relations are naturally formulated in this way~\cite{Gallavotti1995,Lebowitz1999,Gingrich2016}, as are theories of fluctuating hydrodynamics~\cite{Bertini2002,Bodineau2004,Bertini2015}.  Large deviations of time-integrated quantities have provided insight into glassy dynamics~\cite{Garrahan2007,Hedges2009,Malins2012}, sheared systems~\cite{Evans2004,Baule2008,JackEvans2016}, active matter~\cite{Grandpre2018,Nemoto2019,mallmin2019,Tociu2019,Keta2021,Fodor2022arcmp,Yan2022}, and other model systems~\cite{Bertini2005,Derrida2007,Baek2017,Appert2008,Hurtado2014,Jack2015}.  The theory applies to rare events where a physical observable undergoes a significant fluctuation that is sustained over a very long time period.  Also, the mechanisms for these events can be characterised by considering biased ensembles of trajectories~\cite{Lecomte2007,Garrahan2009,Jack2020ergo}, whose construction is inspired by the canonical ensemble of equilibrium statistical mechanics.  

Biased ensembles may support dynamical phase transitions~\cite{Bertini2005,Derrida2007,Garrahan2007,Appert2008,Baek2017}, often associated with spontaneous symmetry breaking.  
The properties of biased ensembles can also be reproduced as the typical dynamics of auxiliary systems, where suitable control forces are added~\cite{Popkov2010,Jack2010,Chetrite2015,Jack2020ergo}: these are sometimes called Doob forces because of theoretical connections to Doob's $h$-transform~\cite{Popkov2010,Chetrite2015}.  Characterisation of such forces is helpful for extracting physical insight from large-deviation computations~\cite{Tociu2019,Keta2021,Fodor2022arcmp}, and for numerical methods~\cite{Nemoto2017first,Nemoto2016,Ray2018,Banuls2019,Rose2021,Yan2022}.  

In this work, we use numerical simulations to analyse a dynamical phase transition in a system of interacting particles, moving in a two-dimensional channel.  The particles follow Langevin dynamics (with inertia); the ensembles are biased by a measurement of particle clustering, which plays a similar role to the dynamical activity considered in previous work~\cite{Hedges2009,Jack2015,Dolezal2019}.  
On biasing to large clustering, the system undergoes a dynamical phase transition and spontaneously breaks symmetry, leading to an increased density near one of the walls of the channel.  

In contrast to exclusion processes and other lattice models, this system can be analysed at a \emph{mechanical} level, including the interparticle forces and the local stress tensor~\cite{irving1950}.  We exploit this approach to gain physical insight into the phase transition.  In the symmetry-broken state, we find  a stress gradient that extends into the bulk of the system; this must be balanced by a suitable body force, to sustain the asymmetric density.  We show that this force comes from thermal noise forces, which develop non-zero average values in response to the bias.  
Physically, this shows that large deviation events with increased clustering involve non-typical instances of the thermal noise forces, which push particles towards the walls of the channel, where they become localized.  
This result reveals the physical origin of the Doob forces, and it also provides a way to measure them directly, which is not possible with standard methods.
We emphasize that our analysis follows directly from the (Newtonian) equations of motion: this mechanical perspective should therefore be applicable to a broad array of rare events and large deviations.

In the following, we define the model and relevant observables in Sec.~\ref{sec:model}.  Results are presented in Sec.~\ref{sec:results}, including the existence of the dynamical phase transition, the behaviour of the stress, and the connection to fluctuating hydrodynamics.  We conclude in Sec.~\ref{sec:outlook} by summarising the consequences of these results, and their relevance for future work.

 \section{Model and methods}
\label{sec:model}

\subsection{System}

We consider $N$ particles with mass $m$ and diameter $l_0$, interacting by a Weeks-Chandler-Andersen (WCA) potential~\cite{Weeks1971} of strength $\epsilon$, in a box of size $L_x\times L$.  It has periodic boundaries in the $y$-direction, and walls at $x=0,L_x$.

Particle $i$ has position $\bm{r}_i=(x_i,y_i)$ and momentum $\bm{p}_i=m\dot{\bm{r}}_i$.  Its equation of motion is
\beq
\dot{\bm{p}}_i 
= -\nabla_i U_{\rm int} - \hat{\bm{x}} V_{\rm w}'(x_i) - \gamma \bm{p}_i + \sqrt{2\gamma mT} \bm{\eta}_i
\label{equ:eom}
\eeq
where $U_{\rm int}$ and $V_{\rm w}$ are the potential energies for particle-particle and particle-wall interactions respectively (see below); also $\hat{\bm{x}}$ is a unit vector in the $x$-direction, $\gamma$ is the frictional damping rate, $T$ is the temperature, and the $\bm{\eta}_i$ are independent unit white noises with
$\langle \eta_i^\mu \eta_j^\nu \rangle= \delta_{ij} \delta^{\mu\nu} \delta(t-t')$, where Greek indices indicate Cartesian components.
To maintain a compact notation, $T$ denotes the thermal energy, that is $T=k_{\rm B}T_{\rm ph}$ where $T_{\rm ph}$ is the physical temperature and $k_{\rm B}$ is Boltzmann's constant.  It is also common to use $\gamma$ to denote a friction constant: our frictional damping force is $m\gamma\dot{\bm{r}}_i$ so the friction constant is $m\gamma$ in our notation.

The interaction 
potential between particle $i$ and the walls is of truncated Lennard-Jones type:
\begin{multline}
V_{\rm w}(x_i) = 4\epsilon \left( \left|\frac{l_0}{x_i-x_{\rm w}}\right|^{12} - \left|\frac{l_0}{x_i-x_{\rm w}}\right|^{6} + \alpha \right)
\\ \times \Theta\big(l_{\rm cut}- |x_i-x_{\rm w}|\big)
\end{multline}
where $\Theta$ is the Heaviside function, $|x_i-x_{\rm w}|$ is the distance from the particle to the nearest wall, the cutoff is $l_{\rm cut}=3l_0/2$, 
and $\alpha$ is chosen to ensure continuity of the potential at the cutoff ($\alpha=0.080$).
We take
$L_x=L + 2\Delta L$, where $\Delta L=l_0(2^{1/6}-(1/4))$ accounts for the excluded volume of the walls.  
Specifically, we vary $N$ and $L$ together, while fixing the (dimensionless) particle concentration $\rhobar = Nl_0^2/L^2$ and also $\Delta L$.  This choice ensures that the \emph{local} particle concentration remains constant in the bulk of the system (away from the walls).

The interaction potential between particles $i$ and $j$ is a Weeks-Chandler-Andersen (WCA) potential~\cite{Weeks1971}:
\beq
U(r_{ij}) = 4\epsilon \left[ \left(\frac{l_0}{r_{ij}}\right)^{12} - \left(\frac{l_0}{r_{ij}}\right)^{6} +\frac14 \right] \Theta\big( l_{\rm WCA}- r_{ij}\big) \; ,
\eeq
where $r_{ij}$ is the distance between the particles and $l_{\rm WCA}=2^{1/6}l_0$ is the usual WCA cutoff.  Hence the potential energy associated with these interactions is
\beq
U_{\rm int} = \sum_{1 \leq i < j \leq N} U(r_{ij}) \; .
\eeq


{In addition to the dimensionless concentration $\rhobar$ defined above and the number of particles $N$, a natural set of dimensionless control parameters of the system is obtained by identifying $v_{\rm 0}=\sqrt{T/m}$ as the thermal velocity, and rescaling all variables in terms of $l_0,v_0,T$.  One obtains two dimensionless parameters:
\beq
\tilde\gamma = \frac{\gamma l_0}{v_0},  \qquad \tilde\epsilon = \frac{\epsilon}{T} \; .
\eeq
The parameter $\tilde\epsilon$ determines the strength of interactions, while
$\tilde\gamma$ determines how strongly damped is the particle motion.}   Since $D_0=T/(m\gamma)$, we can also write
$\tilde\gamma = (l_0v_0/D_0)$, so the damping determines the ratio of the thermal velocity to the diffusion constant (as usual).
The simulation time step is $\delta t$, it is fixed by taking the dimensionless parameter $(v_0 \delta t / l_0) = 0.002$.

Throughout this work we take $\rhobar=0.48$, a moderate density where particles interactions are significant, but the system is not crowded enough to cause slow dynamics or crystallisation.  We set $\tilde\epsilon=1$ (we expect results to depend weakly on this parameter) and $\tilde\gamma=10$, which corresponds to strong damping.
Previous work on biased ensembles focussed on the overdamped limit.  Here we consider finite damping so that momenta are well-defined, this is convenient for the mechanical analysis.  However, for the large $
\tilde\gamma$ that we consider, we expect the physical behaviour to be similar to the overdamped limit.
When presenting numerical results, we use non-dimensional units in which $l_0,m,T$ are set to unity (so $v_0=1$ also).

\subsection{Time-integrated clustering, and biased ensembles}

We analyse dynamical trajectories over an observation time $t_{\rm obs}$.  Our results are controlled by the hydrodynamic behavior of the system, so we define a dimensionless observation time $\tau_{\rm obs}=t_{\rm obs}/\tau_L$ where $\tau_L = L^2/D_0$ is the hydrodynamic time scale, in which $D_0=T/(m\gamma)$ is the (bare) particle diffusivity.
To measure clustering between particles, write $r_{ij}=|\bm{r}_i-\bm{r}_j|$ 
and define 
\beq
Q(r_{ij}) = \begin{cases} 
l_{\rm WCA}/l_0 , & r_{ij} < (l_{\rm WCA}/2) 
\\
2(l_{\rm WCA} - r_{ij})/l_0 , & (l_{\rm WCA}/2) < r_{ij} < l_{\rm WCA}
\\ 
0 , & r_{ij} > l_{\rm WCA} 
\; .
\end{cases}
\eeq
This function interpolates between a value of order unity when particles $i$ and $j$ are very close, and zero when their distance exceeds the cutoff distance $l_{\rm WCA}$.
The total  clustering within a trajectory is measured by integrating over time and summing over all pairs of particles:
\beq
C(\tau_{\rm obs}) = \frac{D_0}{L^2} \int_0^{t_{\rm obs}}  \sum_{1\leq i< j \leq N} Q(r_{ij}(t))  dt \; .
\label{equ:C-def}
\eeq
This quantity has been rendered dimensionless through normalisation by the hydrodynamic time scale.

\begin{figure}
\includegraphics[width=8cm]{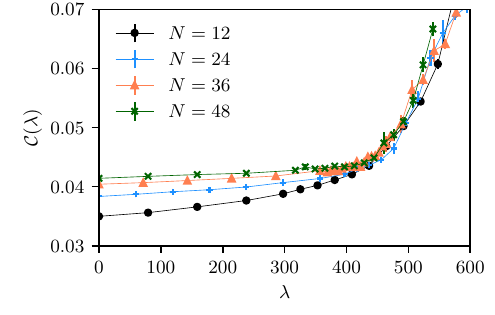}
\caption{%
Average {(dimensionless)} clustering ${\cal C}(\lambda)$ for the biased ensemble, for several system sizes.  There is a change in behavior at $\lambda=\lambda_c\approx 450$, which corresponds to a dynamical phase transition.}
\label{fig:OP}
\end{figure}

We consider biased ensembles 
of trajectories 
in which the average of any observable quantity $A$ is given by
\beq
\langle A \rangle_\lambda = \frac{ \langle A {\rm e}^{\lambda C(\tau_{\rm obs})} \rangle_0 }{  \langle {\rm e}^{\lambda C(\tau_{\rm obs})} \rangle_0 }
\label{equ:bias}
\eeq
where $\lambda$ is a {dimensionless} biasing parameter and $\langle \cdot \rangle_0$ represents an average in the equilibrium state of the system.   To understand the physical meaning of the bias, consider the average {(dimensionless)} clustering per particle in this ensemble:
\beq
{\cal C}(\lambda) = \frac{1}{N\tau_{\rm obs}}\langle C(\tau_{\rm obs})\rangle_{\lambda} \; .
\eeq
For large $\tau_{\rm obs}$, the biased ensembles reproduce the mechanism for (rare) fluctuations of $C(\tau_{\rm obs})$~\cite{Lecomte2007,Garrahan2009,Chetrite2015,Jack2020ergo}.  The value of $\lambda$ encodes the size of the fluctuation, according to
 $C(\tau_{\rm obs})\approx N\tau_{\rm obs} {\cal C}(\lambda)$.  Positive $\lambda$ corresponds to increased clustering.
The factor of $L^2$ in (\ref{equ:C-def}) means that $\lambda$ biases the hydrodynamic behavior, but has a weak effect on individual particle dynamics~\cite{Jack2020ergo,Dolezal2019}, see Sec.~\ref{sec:mft} for further details. 
Comparing with previous work for ensembles biased by the dynamical activity~\cite{Garrahan2009,Jack2015}, we expect large clustering to correspond to low dynamical activity~\cite{Fullerton2013}.  

\section{Results}
\label{sec:results}

\subsection{Dynamical phase transition}
\label{sec:phase-transition}

\begin{figure}
\includegraphics[width=80mm]{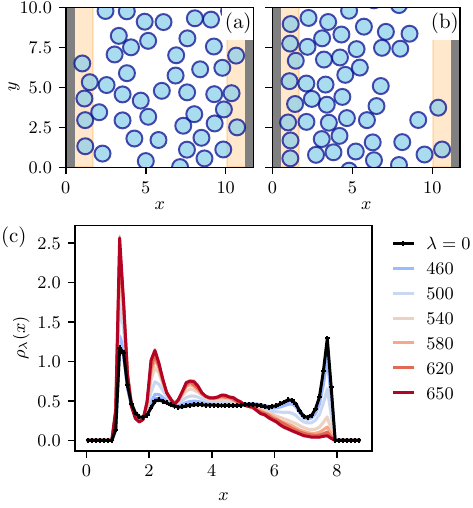}
\caption{Representative snapshots of a system with $N=48$ particles. (a)~Equilibrium state $\lambda=0$. (b)~Biased state $\lambda=870$ with broken symmetry.  The orange shaded regions indicate the range of the wall potential $V_{\rm w}$. (c)~Average density in biased ensembles for $N=24$, showing symmetry breaking, and layering of particles at the left wall.  Statistical uncertainties are comparable with line widths.  {Recall that distances are measured throughout in units of the particle diameter $l_0$.}
}
\label{fig:rho}
\end{figure}

We use transition path sampling (TPS)~\cite{Bolhuis2002} to sample these biased ensembles.  We take $\tau_{\rm obs}=0.252$, which is sufficiently large to reveal the phase transition, as we will see below.  As in~\cite{Dolezal2019}, we include auxiliary forces in the dynamics to improve sampling, similar to~\cite{Nemoto2016}.
Fig \ref{fig:OP} illustrates the behaviour of ${\cal C}(\lambda)$ as $\lambda$ is increased from zero, for
several system sizes, with fixed $\rhobar,\tau_{\rm obs}$.  There is a change in behaviour for $\lambda= \lambda_c\approx450$, which
becomes increasingly pronounced for larger systems.  This is a dynamical phase transition where symmetry is spontaneously broken (see Fig.~\ref{fig:rho}, discussed below).
The large numerical value of $\lambda_c$ is attributable to the small values of ${\cal C}$ in the unbiased system; the steepness of the WCA potential means that particle separations are never much less than $l_{\rm WCA}$.

Fig.~\ref{fig:rho} illustrates the spontaneous symmetry breaking for $\lambda>\lambda_c$: particles accumulate near one wall of the system.  This can be be quantified by
the average local density, defined as $\rho_\lambda(\bm{r})=\sum_i \langle \delta(\bm{r}-\bm{r}_i) \rangle_\lambda$. 
Accumulation at either wall is equally likely: as in equilibrium phase transitions, it is convenient to break the symmetry in the numerical computation, to obtain a clear signature of the symmetry-broken state.  This is achieved by using asymmetric auxiliary forces in the TPS method, as discussed in Sec.~\ref{sec:control}, below. 

The main features of this transition can be explained within macroscopic fluctuation theory (MFT)~\cite{Bertini2015}.  Following~\cite{Lecomte2012,Jack2015,Dolezal2019,Dolezal2021}, one assumes that the clustering is a function of the local density, in which case the hydrodynamic behavior of biased ensembles can be predicted by minimization of a dynamical action that depends on the density alone.  This indicates that (i) ${\cal C}(\lambda)$ should approach a scaling function in the limit of large system size $N$ (consistent with Fig.~\ref{fig:OP}); and (ii) the density profile should respond most strongly on the largest relevant length scale, which is the system size.  This is consistent with Fig.~\ref{fig:rho} (and with Fig.~\ref{fig:stress}, below).  See Sec.~\ref{sec:mft} for additional information on this hydrodynamic analysis.

\subsection{Stress and force balance}

So far, the analysis of this dynamical phase transition has similar features to previous work on simpler systems~\cite{Lecomte2012,Baek2017,Dolezal2019}.  We now focus on mechanical properties of this biased ensemble, which have not been considered before, to our knowledge.  
We write the equation of motion \eqref{equ:eom} as
\beq
\dot{\bm{p}}_i  +\nabla_i U_{\rm int} = \bm{f}_i
\label{equ:dot-pi}
\eeq
where
\beq
\bm{f}_i =  - \hat{\bm{x}} V_{\rm w}'(x_i) - \gamma \bm{p}_i + \sqrt{2\gamma mT}  \bm{\eta}_i \; .
\label{equ:body-f}
\eeq
is the \emph{body force} on particle $i$.  These are forces that do not conserve the total particle momentum $\bm{P}=\sum_i \bm{p}_i$.
{[In fact, $\dot{\bm{P}} = \sum_i \bm{f}_i$, which follows because $(\nabla_i + \nabla_j) U(r_{\rm ij})=0$.  Physically: the interparticle forces obey Newton's 3rd law so they can't change the total particle momentum, but the body forces can inject momentum from the particles' environment.]}

By analogy with $\rho_\lambda$, define the average
 body-force density as $\bm{F}_\lambda(\bm{r}) = \sum_i \langle \bm{f}_i \delta(\bm{r}-\bm{r}_i) \rangle_\lambda$.  Cartesian components of $\bm{F}_\lambda$ are denoted by ${F}^\alpha_\lambda$ with $\alpha\in\{x,y\}$, we use a similar notation for other vectors and tensors.
The biased ensembles of trajectories that we consider are homogeneous in time, which means that they 
must have balanced mechanical forces.  Specifically, body forces must be balanced by stress gradients:
\beq
 F^\alpha_\lambda(\bm{r})  = - \sum_{\beta} \nabla_\beta \Pi_\lambda^{\alpha\beta}(\bm{r})
 \label{equ:mech-balance}
\eeq
where $\Pi_\lambda$ is the local stress tensor (a position-dependent $2\times2$ matrix),
and $\nabla_\beta$ indicates differentiation with respect to $r^\beta$.
We use the procedure of Irving and Kirkwood~\cite{irving1950} to compute the stress in the biased ensemble, see Appendix~\ref{app:IK} for details.
Within that framework, we emphasise that \eqref{equ:mech-balance} can be derived directly from the equations of motion of the system, there is no assumption that the system be near-equilibrium, only that it is stationary.
Hence the formalism also applies in the biased ensemble.

\begin{figure}
\includegraphics[width=8cm]{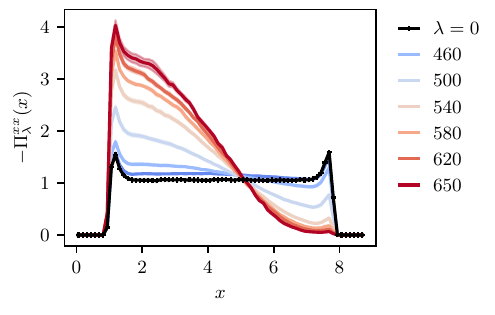}
\caption{Stress tensor (specifically, $-\Pi^{xx}_\lambda$) for the same biased ensembles as Fig.~\ref{fig:rho}(b) with $N=24$.  
The bias generates a stress gradient that extends into the bulk.  The shading indicates the statistical uncertainties (standard error).
See Fig.~\ref{fig:N48} for similar results with $N=48$.
{(Consistent with our convention, the unit of stress is $T/l_0^3$ in all numerical measurements.)}
}
\label{fig:stress}
\end{figure}

The biased states retain translational invariance along the $y$-direction, 
so (\ref{equ:mech-balance}) reduces to
\beq
F^x_\lambda(x) = - \nabla_x \Pi^{xx}_\lambda(x) \; .
\label{equ:balance}
\eeq
For the unbiased dynamics, Eq~\eqref{equ:body-f} implies that
$
 F^x_0(x)  = - V'_{\rm w}(x)  \rho_0(x)  
$.
Away from the wall, this quantity vanishes, so $\Pi^{xx}_0$ is independent of $x$ [by (\ref{equ:balance})]: in fact $-\Pi^{xx}_0$ equals the pressure of the bulk fluid.
This situation is shown in black in Fig.~\ref{fig:stress}: the stress is constant in the bulk, while body forces generate gradients near the wall.

 The biased ensemble is time-reversal symmetric which means that $\langle \bm{p}_i \delta(\bm{r}-\bm{r}_i)\rangle_\lambda=0$, so the friction term in (\ref{equ:body-f}) does not contribute to the average body force.  Hence,
\beq
F^x_\lambda(x) = - V'_{\rm w}(x) \rho_\lambda(x) +  \sum_i \sqrt{2\gamma mT} \langle  {\eta}_i^x \delta(x-x_i)\rangle_\lambda \; .
\label{equ:F-eta}
\eeq 
For $\lambda>\lambdaC$, one sees from Fig.~\ref{fig:stress} that there is a stress gradient that extends into the bulk of the channel, so Eq.~\ref{equ:balance} requires a non-zero body force $F^x_\lambda$ there.  But $V'_{\rm w}=0$ in the bulk, so a non-zero body-force in \eqref{equ:F-eta} requires that \emph{the averaged Langevin noises must be non-zero}, in the biased ensemble.
This result may be counter-intuitive: it occurs because the biased ensemble changes the probabilities of different trajectories, so it also changes the probabilities of particular realisations of the thermal forces $\bm{\eta}_i$.  This leads to a position-dependent body force within the biased ensemble.  Indeed, some such force must be present, to maintain the asymmetric density profiles in Fig.~\ref{fig:rho}.

\subsection{Optimal-control representation and Doob's transform}
\label{sec:control}

\newcommand{\bteta}{\tilde{\bm{\eta}}}

In fact, these body forces are already familiar from theories of biased ensembles.  
As described above, the effects of the bias $\lambda$ can be reproduced by adding control forces to the equations of motion, leading to an auxiliary model~\cite{Jack2010,Chetrite2015,Jack2020ergo} as in Doob's transform.  The equation of motion for the auxiliary model is
\beq
\dot{\bm{p}}_i = 
 -\nabla_i U_{\rm int} - \hat{\bm{x}} V_{\rm w}'(x_i) - \gamma \bm{p}_i + \bm{\phi}_i + \sqrt{2\gamma mT}  \bteta_i
\label{equ:eom-aux}
\eeq 
where $\bm{\phi}_i$ is a control force on particle $i$ that depends (in general) on the positions and momenta of all particles in the system, and $\bteta_i$ is a white noise with the same statistical properties as $\bm{\eta}_i$.  Denote averages in the unbiased steady state of this auxiliary model by 
$\langle  A \rangle_{\rm aux}$.
Then the Doob transform yields~\cite{Jack2010,Chetrite2015} that 
\beq
\langle A \rangle_{\lambda} =\langle A \rangle_{\rm aux} 
\label{equ:AA}
\eeq
for a large class of observables $A$, in the limit of large $t_{\rm obs}$. The control forces $\bm{\phi}_i$ depend on $\lambda$, so $\langle A \rangle_{\rm aux}$ does too.  {Note that the auxiliary model does not capture transient behavior of the biased ensemble for times $t\approx 0$ and $t\approx t_{\rm obs}$~\cite{Garrahan2009,Chetrite2015}.  This restricts the the class of observables $A$ to those for which transient effects have a negligible contribution for large $t_{\rm obs}$.
All observables considered here are within this class.}

In the steady state of the auxiliary model, the $\bm{\phi}_i$ act as body forces.  The stress tensors of the biased ensemble and the auxiliary model are equal [by \eqref{equ:AA}] which means that their body forces must also be equal [by (\ref{equ:mech-balance})].  
Equating these forces yields
\beq
 \sum_i \langle \bm{\phi}_i \delta(\bm{r}-\bm{r}_i)\rangle_{\rm aux}
=
\sum_i\sqrt{2\gamma mT}\langle  \bm{\eta}_i \delta(\bm{r}-\bm{r}_i)\rangle_\lambda \; ,
\label{equ:eta-phi}
\eeq
where we used that $\langle\bteta_i\delta(\bm{r}-\bm{r}_i)\rangle_{\rm aux}=0$.  
The left hand side of \eqref{equ:eta-phi} is the contribution of the $\bm{\phi}_i$ to the body force of the auxiliary model,
and
the right hand side is the contribution of the noise forces to the body force in the biased ensemble.
We denote the left hand 
side of (\ref{equ:eta-phi}) by $\rho_\lambda(\bm{r}) \bm{\phi}_{\rm ave}(\bm{r})$, so that $ \bm{\phi}_{\rm ave}(\bm{r})$ is the average control force on a particle at $\bm{r}$.

\begin{figure}
\includegraphics[width=8cm]{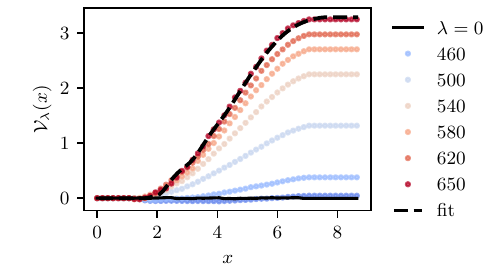}
\caption{The Doob stress ${\cal V}_\lambda$, for the biased ensembles of Fig.~\ref{fig:stress}.  The gradient of this stress corresponds to a force which acts to the left.  The fitted function (for the largest $\lambda$) is discussed in main text.}
\label{fig:doob}
\end{figure}

Eq.~\eqref{equ:eta-phi} means that the \emph{control forces that appear in Doob's transform have a mechanical interpretation}: they generate the biased noise forces identified in (\ref{equ:F-eta}).  
To infer the control forces themselves, 
combine (\ref{equ:mech-balance},\ref{equ:F-eta},\ref{equ:eta-phi}) to obtain
\beq
\rho_\lambda(x) {\phi}_{\rm ave}^x(x) = V'_{\rm w}(x) \rho_\lambda(x) - \nabla_x \Pi^{xx}_\lambda(x) \; .
 \label{equ:eta-Pi}
\eeq
That is, the (averaged) control force can be estimated from measurements of the density $\rho_\lambda$ and the stress $\Pi_\lambda$, within the biased ensemble.  

Of course, the Doob force $\bm{\phi}_i$ may depend on the positions of all particles, while measuring the stress gradient only gives the average force $ \bm{\phi}_{\rm ave}(\bm{r})$.  Eq.~(\ref{equ:eta-phi}) could be generalised by adding additional delta functions inside the averages, such that Doob forces can be related to  conditional averages of $\bm{\eta}_i$.  However, we concentrate here on the information that is available from the stress tensor, since this is more easily measured than the noise forces themselves.

Integrating (\ref{equ:eta-Pi}), we express the control force as ${\phi}_{\rm ave}^x(x)  = -{\cal V}_\lambda'(x)/\rho_\lambda(x)$ 
where we define the \emph{Doob stress},
\beq
{\cal V}_\lambda(x) = \Pi^{xx}_\lambda(x) -  \int_0^x V'_{\rm w}(u) \rho_\lambda(u) du \; .
\eeq
This is the part of the stress that originates from the noise forces.  
It is estimated numerically in Fig.~\ref{fig:doob}.  To approximate $\bm{\phi}$, we fit the stress as 
${\cal V}_\lambda(x) \approx -\int_0^x \rho_\lambda(u) \phi_{\rm est}(u) du$, using 
\beq
\phi_{\rm est}^x(x) = \begin{cases} -a-bx ,  & l_{\rm cut} < x < L_x - l_{\rm cut}
\\
0 , & \hbox{otherwise.}  \end{cases}
\eeq
This function is zero near the walls, but negative in the bulk (corresponding to a force towards the left wall). 
Close inspection of Fig.~\figVBE\ shows that the resulting fit to the stress is not perfect in the regions close to the wall, where layering takes place.  However, it does capture the behaviour in bulk, and hence the hydrodynamic response to the bias.

A sample fit is shown in Fig.~\ref{fig:doob}.  Since the stress gradient occurs on the scale of the system size $L$, this control force is of order $1/L$.  This is another indication of the hydrodynamic response to the bias, which appears when a large number of particles each receives a weak bias, leading to a macroscopic response~\cite{Jack2020ergo}.

The estimated control force $\phi_{\rm est}$ can be inserted into (\ref{equ:eom-aux}) as $\bm{\phi}_i=\hat{\bm{x}}\phi_{\rm est}(x_i) $, where $\hat{\bm{x}}$ is a unit vector in the $x$-direction.  This gives an auxiliary model whose density profile is similar to the biased ensemble -- this is not the exact Doob dynamics but it can be used within the TPS algorithm, to propose new trajectories that are more likely to be accepted.  Using this method and systematically refining $\phi_{\rm est}$ greatly improves numerical performance of the algorithm, similar to~\cite{Ray2018,Dolezal2019,Nemoto2016,Banuls2019,Yan2022}.  Exactly this method was used to obtain the data shown here -- without such control forces, the simulations for larger systems and larger biases would have been computationally intractable.
In this sense, insights from the mechanical analysis can be exploited to improve numerical methods.

 {Note that including control forces within TPS does not change the result of the numerical computation, which always samples the ensemble defined by (\ref{equ:bias}), as long as $t_{\rm obs}$ is large enough: this feature was discussed extensively in~\cite{Nemoto2016} in the context of population dynamics algorithms, and also in \cite{Dolezal2019} for TPS.  To understand it, one should think of TPS as a Monte Carlo (MC) procedure for trajectories of the system.  Including auxiliary forces amounts to a different set of proposed MC updates, which does not change the ensemble  being sampled [Eq.~\eqref{equ:bias}], but may improve the acceptance rate.  To verify that the target ensemble is not affected by the control forces, we checked that for $\lambda<\lambda_c$ then the sampled density profile is symmetric, even when asymmetric auxiliary forces are used.   For $\lambda>\lambda_c$, the auxiliary forces still do not affect the sampled distribution, but they do improve the acceptance rate within the broken-symmetry state.  (Without such forces, the TPS algorithm tends to propose trajectory updates where the particles move away from the walls.  These updates tend to be rejected, leading to inefficient sampling.)}

\subsection{Hydrodynamic response in biased ensemble}
\label{sec:mft}

   \begin{figure*}
\includegraphics[width=160mm]{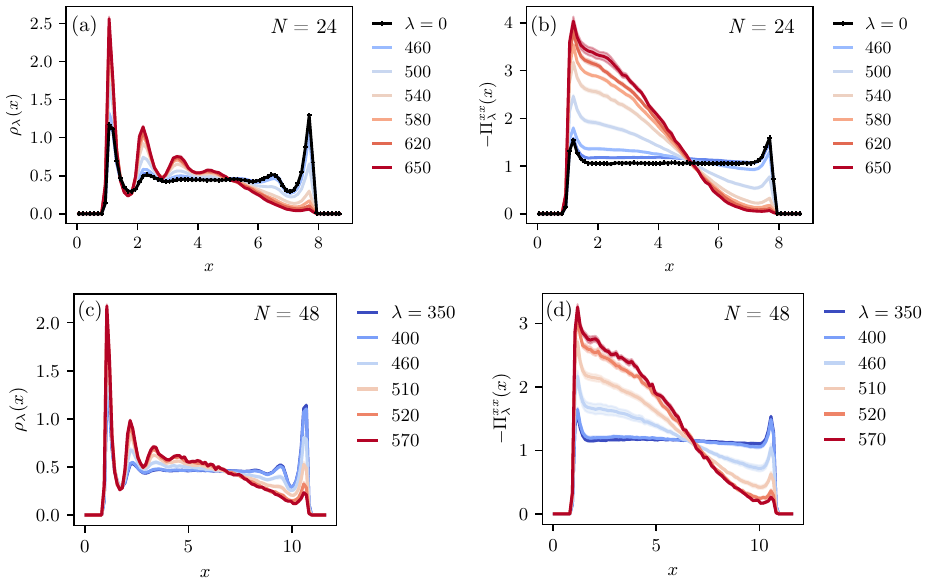}
 \caption{Density and stress profiles for $N=24$ and $N=48$.  
 The top panels are repeated from Figs.~\figRho\ and \figStress\ of the main text.  The system sizes are $L_x=8.82l_0$ and $11.74l_0$ (2 decimal places).}
 \label{fig:N48}
 \end{figure*}

We noted in Sec.~\ref{sec:phase-transition} that the field $\lambda$ biases the hydrodynamic behaviour of the system.  This section gives some extra information on this point.
 
Based on the analogy between biased ensembles and thermodynamics~\citeBias, it is natural to define a measure of clustering that is extensive in space and time:
\beq
{\cal Q}(t_{\rm obs}) = \int_0^{t_{\rm obs}} \sum_{i<j} Q_{ij}(r_{ij}(t)) dt \; .
\eeq
Then the definition of the biased ensemble in (\eqDefBias) is
\beq
\langle A \rangle_\lambda = \frac
{ \left\langle A \exp\!\left[ \frac{\lambda D_0}{L^2} {\cal Q}(t_{\rm obs}) \right] \right\rangle_0 }
{ \left\langle \exp\!\left[ \frac{\lambda D_0}{L^2} {\cal Q}(t_{\rm obs}) \right] \right\rangle_0 } \; .
\label{equ:bias-hydro}
\eeq

In systems without hydrodynamic modes, it would be natural to use a definition similar to (\ref{equ:bias-hydro}), but with $\lambda D_0/L^2$ replaced by an intensive field that  is often denoted by $s$.  Then one would keep $s$ fixed in a joint limit $L,t_{\rm obs}\to\infty$, which corresponds to applying a bias of order unity to each particle, even as the system size tends to infinity.

In the present context, we fix $\lambda$ as $L,t_{\rm obs}\to\infty$.  Hence one sees from (\ref{equ:bias-hydro}) that the intensive field $s=\lambda D_0/L^2$ is being reduced as the system size increases.  The situation is familiar in systems with hydrodynamic modes~\cite{Appert2008,Lecomte2012,Dolezal2019}, where weak biases on many particles can add up coherently to generate strong responses.  This lends a degree of universality to such behaviour, which can be captured by theories like macroscopic fluctuation theory~\citeMFT.  Similar mechanisms are at work in the system considered here.

We sketch the MFT analysis of the transition (following the analysis of~\citeJakubMany\ and \cite{Lecomte2012,Jack2015} for one-dimensional systems): Rescale in hydrodynamic units as $\tilde{\bm{r}}=\bm{r}/L$ and $\tilde{t}=D_0t/L^2$ and assume that the clustering can be parameterised in terms of the local density as ${\cal Q}(\tilde{t}_{\rm obs}) \approx \int_0^{\tilde{t}_{\rm obs}} \int q(\rho(\tilde{\bm{r}},\tilde{t})) d\tilde{\bm{r}} d\tilde t$ for some ``clustering density'' ${q}(\rho)$.  Then the MFT action for the biased ensemble would be
\beq
S[\rho,\tilde J] = \int \left[ \frac{|\tilde J + D(\rho) \nabla \rho |^2}{4 \sigma(\rho)} - \lambda {q}(\rho) \right] d\tilde{\bm{r}} d\tilde t
\eeq
where $\tilde J$ is the hydrodynamic current that obeys $\partial_{\tilde t} \rho = -\tilde\nabla \cdot \tilde J$, and $D,\sigma$ are the density-dependent  diffusivity and mobility of MFT.  (The effect of the walls would be incorporated through hard boundaries, after rescaling to the hydrodynamic scale.)  Using time-translation invariance and time-reversal invariance of the biased ensemble, the time integral can be ignored and one has $\tilde J=0$.   Finally, the hydrodynamic density would be obtained by minimising the functional
\beq
{\cal L}[\rho] = \int \left[ \frac{| D(\rho) \nabla \rho |^2}{4 \sigma(\rho)} - \lambda {q}(\rho) \right] d\tilde{\bm{r}} \; .
\eeq
 In the present context, the functional forms of $D,\sigma,{q}$ are not known, so a quantitative analysis is not possible.  However, it is clear on general grounds that for ${q}''>0$, positive $\lambda$ will drive the system to an inhomogeneous state, similar to transitions in other contexts~\cite{Lecomte2012,Dolezal2019}.

 Fig.~\ref{fig:N48} shows results for $N=48$, compared with the corresponding results for $N=24$, as already shown in Figs.~\ref{fig:rho} and~\ref{fig:stress}.
 Ignoring the layering effects near the walls (which are non-hydrodynamic in nature), and focussing on length scales of the order of the system size, one sees a semi-quantitative match of the density and stress gradients that develop as a function of $\lambda$.
For full consistency with MFT, one would require that the density and stress would be scaling functions of $x/L$, in the limit of large systems $N\to\infty$.  In that case, the stress gradient in bulk would be $O(1/L)$, as discussed in the main text.  

In practice, the systems considered are far from the limit $N\to\infty$, so microscopic length scales also affect the results.  Still, the observed behaviour is consistent with the hydrodynamic theory.

\section{Outlook}
\label{sec:outlook}

{We summarize the results of this work: These two-dimensional systems of interacting particles support a dynamical phase transition, associated with large deviations where the particles are clustered more than usual.   At this transition, a symmetry is spontaneously broken, leading to particle aggregation at a wall.  These transitions are naturally studied via biased ensembles of trajectories, which we have implemented numerically by TPS.  In addition, while some similar behaviour is observed in lattice models~\cite{Lecomte2012,Baek2017}, the fact that this system follows Newtonian dynamics means that the balance of mechanical forces can be investigated within the biased ensemble.  In particular, the stress tensor of Irving and Kirkwood~\cite{irving1950} can be computed within the biased ensemble, and yields useful information, even in systems far from equilibrium.   (This property distinguishes the stress from objects like the thermodynamic free energy, which is no longer meaningful in biased ensembles.)

This mechanical analysis shows that if particle motion depends on the bias parameter $\lambda$, this dependence must have its origin in some underlying forces.
%
These forces appear as non-zero averages for the Langevin noises $\bm{\eta}_i$, and as control forces that appear in Doob's transform.  Their presence can be inferred from the stress tensor.  
The physical picture is that the biased ensemble changes the probabilities of trajectories of the system, which changes the statistical properties of the noise.   In particular, the bias tends to selects noise realisations that push particles towards one of the walls.  This is the cause of the particle aggregation at the walls, and of the broken symmetry.}

We emphasise that this mechanical perspective is not specific to the system analysed here, it is valid in any system where the  stress tensor can be defined.  In particular, we have considered inertial motion, but the same methodology is applicable in the overdamped limit, so large deviations of diffusions~\cite{Chetrite2015} can be analysed in the same way.  For these reasons, we hope that this approach will provide new opportunities for understanding large deviations in physical systems, and for characterising them numerically.

\begin{acknowledgments}
We thank Kris Thijssen, Mike Cates, and Tal Agranov for helpful discussions.  We are grateful to the EPSRC for support through a studentship for JD (ref EP/N509620/1) and research funding to RLJ  (ref EP/T031247/1).
{The data supporting this publication are available at https://doi.org/10.17863/CAM.86077.}
\end{acknowledgments}

\begin{appendix}

\section{Irving-Kirkwood stress}
\label{app:IK}

\subsection{Definition, and derivation of (\eqBal)}

This section reviews the computation of the IK stress tensor~\citeIK. 
For a modern presentation, we follow~\citeParola\ (in particular, the Supplemental Material of that work).

As a preliminary for computing the stress, we consider the particle current.
Define the empirical particle density and momentum density as
\begin{equation}
\hat\rho(\bm{r}) = \sum_i \delta(\bm{r} - \bm{r}_i ) , \qquad \hat{\bm{p}}(\bm{r}) = \sum_i \delta(\bm{r} - \bm{r}_i ) \bm{p}_i
\; .
\end{equation}
These quantities have implicit time-dependence via the positions and momenta.
Then the time derivative of the particle density is 
\beq
\partial_t {\hat\rho}(\bm{r}) = - \sum_i (\bm{p}_i/m) \cdot \nabla \delta(\bm{r} - \bm{r}_i ) 
\; ,
\label{equ:dot-rho}
\eeq
where we used the chain rule and $\dot{\bm{r}}_i = \bm{p}_i/m$ (gradients $\nabla$ are with respect to $\bm{r}$).
Since all particles have equal mass, the particle current may be identified as $\hat{\bm{\jmath}}(\bm{r}) = (1/m) \hat{\bm{p}}(\bm{r})$, and one recognises (\ref{equ:dot-rho}) as the continuity equation
\beq
\partial_t {\hat\rho}(\bm{r}) = -\nabla\cdot\hat{\bm{\jmath}}(\bm{r}) \; .
\eeq

\newcommand{\ff}{\mathcal{F}}

The manipulations so far are familiar from analysis of particle currents.
The IK stress~\citeIK\ is derived by applying similar procedures to the momentum density.  The time derivative of $\hat{\bm{p}}$ is available by using the chain rule together with (\ref{equ:dot-pi}): denoting Cartesian components by Greek indices we obtain
\begin{multline}
\partial_t {\hat p}^\mu(\bm{r}) = \sum_i  f_i^\mu  \delta(\bm{r}-\bm{r}_i) 
+\sum_{i\neq j}  \ff_{ij}^\mu \delta(\bm{r}-\bm{r}_i) 
\\ - \sum_\nu \nabla^\nu \left[ \sum_i \frac{p_i^\mu {p}_i^\nu}{m} \delta(\bm{r} - \bm{r}_i ) \right] 
\label{equ:p-bal}
\end{multline}
where $\bm{\ff}_{ij}$ is the force on particle $i$ from particle $j$, whose components are
\beq
\ff_{ij}^\mu=\frac{24\epsilon(r_i^\mu-r_j^\mu)}{r_{ij} l_0}   \left[ 2\left(\frac{l_0}{r_{ij}}\right)^{13} - \left(\frac{l_0}{r_{ij}}\right)^{7}  \right] \Theta\big( l_{\rm WCA}- r_{ij} \big).
\label{equ:f-wca}
\eeq
Eq.~(\ref{equ:p-bal}) for the momentum is analogous to  Eq.~(\ref{equ:dot-rho}) for the density.
The analogue of the current in this case will be the stress tensor.  To see this, use we use 
the insight of~\citeIK: our discussion follows~\citeParola.  The key point is that (\ref{equ:p-bal}) can be rewritten as
\begin{equation}
\partial_t  {\hat p}^\mu(\bm{r}) = \sum_i  f_i^\mu  \delta(\bm{r}-\bm{r}_i)  + \nabla^\nu \hat\Pi^{\mu\nu}(\bm{r}) \; ,
\label{equ:IK-pdot}
\end{equation}
where we use (throughout this Appendix, but not in the main text) the convention of implicit summation of repeated Greek indices, 
and the IK stress tensor is 
\beq
\hat\Pi^{\mu\nu}(\bm{r}) =  \frac12 \sum_{i\neq j} ({r}_j^\nu - {r}_i^\nu ) h(\bm{r};\bm{r}_i,\bm{r}_j)   \ff_{ij}^\mu - \sum_i \frac{p_i^\mu {p}_i^\nu}{m} \delta(\bm{r} - \bm{r}_i )
\label{equ:IK-stress}
\eeq
in which 
\beq
h(\bm{r};\bm{r}_i,\bm{r}_j) =\int_0^1 \delta\!\left[ \bm{r} - \lambda \bm{r}_j - (1-\lambda) \bm{r}_i\right] d\lambda \; 
\eeq
is a function which distributes unit weight over a straight line connecting $\bm{r}_i$ and $\bm{r}_j$.  The consistency of  (\ref{equ:IK-pdot},\ref{equ:IK-stress}) with (\ref{equ:p-bal}) is demonstrated in App.~\ref{sec:ik-lemma}, below.

Observe that the WCA interaction is always repulsive [the object in square brackets in (\ref{equ:f-wca}) is positive], which means that diagonal elements of $\hat\Pi$ are always negative.  This ensures that the (instantaneous local) mechanical pressure  $\operatorname{tr}(-\hat\Pi/2)$ is always positive.

We now derive (\eqBal) of the main text.
Note that (\ref{equ:IK-pdot}) is an exact identity: it was derived from the equations of motion and it holds for every trajectory of the system.  This means that it can be used to analyse force balance in the biased ensemble.  In particular, we
take the average of (\ref{equ:IK-pdot}) within the biased ensemble, to obtain
\begin{equation}
\langle \partial_t  {\hat p}^\mu(\bm{r}) \rangle_\lambda = \sum_i  \langle  f_i^\mu  \delta(\bm{r}-\bm{r}_i) \rangle_\lambda + \nabla^\nu \langle \hat\Pi^{\mu\nu}(\bm{r})\rangle_\lambda \; .
\label{equ:pdot-ave}
\eeq
The left hand side is zero (we again exclude transient regimes for $t\approx 0,t_{\rm obs})$.  Define also the ensemble-averaged stress tensor as
\beq
\Pi^{\mu\nu}_\lambda(\bm{r}) = \big\langle \hat\Pi^{\mu\nu}(\bm{r}) \big\rangle_\lambda
\;. 
\eeq
Then (\ref{equ:pdot-ave}) becomes $0=F^\mu(\bm{r}) + \nabla^\nu \Pi^{\mu\nu}_\lambda(\bm{r})$ which is (\eqBal) of the main text.  [We used the definition of the body-force density $\bm{F}_\lambda = \sum_i  \langle  \bm{f}_i  \delta(\bm{r}-\bm{r}_i) \rangle_\lambda$.]

\subsection{Validation of the IK stress formula (\ref{equ:IK-stress})}
\label{sec:ik-lemma}

To see that (\ref{equ:IK-pdot}) is equivalent to (\ref{equ:p-bal}), 
we follow~\cite{parola2019,Schofield1982} and note the following property of $h$:
\beq
 (\bm{a}-\bm{b}) \cdot \nabla h(\bm{r};\bm{a},\bm{b}) = \delta(\bm{r}-\bm{b}) - \delta(\bm{r}-\bm{a}) 
\label{equ:magic}
\eeq
where the gradient is with respect to $\bm{r}$.  
To show this:  
Observe that for any path from $\bm{a}$ to $\bm{b}$ and any function $f$ we have
$
f(\bm{b}) - f(\bm{a})  = \int_{\bm{a}\to\bm{b}} \nabla f(\bm{x}) \cdot d\bm{x} 
$.
Considering the straight line path from $\bm{a}$ to $\bm{b}$ and parameterising by $\lambda$, we define
$\bm{r}_\lambda =  \lambda\bm{b} + (1-\lambda)\bm{a}$.  Then
\begin{align}
f(\bm{b}) - f(\bm{a})
& = \int _0^1  \nabla f\big( \bm{r}_\lambda \big) \cdot (\bm{b}-\bm{a})  d\lambda
\nonumber\\
& = \int _0^1 \left[ - \int f(\bm{r}) \nabla \delta\big( \bm{r} - \bm{r}_\lambda\big) d\bm{r}\right] \cdot (\bm{b}-\bm{a})  d\lambda
\end{align}
where the second line uses $\nabla f(\bm{x})  = - \int f(\bm{r}) \nabla \delta(\bm{r}-\bm{x}) d\bm{r}$ (the integral runs over all space, so an integration by parts shows that this applies for any function $f$).  
Using again the definitions of the delta function and of $h$, this can be rearranged as:
\begin{multline}
\int  \left[ \delta(\bm{r}-\bm{b}) - \delta(\bm{r}-\bm{a})
 - (\bm{a}-\bm{b}) \cdot\nabla h (\bm{r};\bm{a},\bm{b}) \right]  
 \\ \times 
 f(\bm{r})   d\bm{r}
 = 0 \; .
\end{multline}
This holds for every function $f$, so the object in square brackets must vanish, this implies (\ref{equ:magic}).

Now take \eqref{equ:magic} with $\bm{a},\bm{b}=\bm{r}_j,\bm{r}_i$ and multiply by ${\cal F}_{ij}^\mu$, yielding
\beq
\nabla^\nu [  ({r}_j^\nu - {r}_i^\nu ) \ff_{ij}^\mu h(\bm{r};\bm{r}_i,\bm{r}_j) ] = \ff_{ij}^\mu \delta(\bm{r}-\bm{r}_i) + \ff_{ji}^\mu \delta(\bm{r}-\bm{r}_j)
\eeq  
(we used that $\ff_{ij}=-\ff_{ji}$).  Using this result with (\ref{equ:IK-stress}), one obtains
\begin{multline}
\nabla^\nu \hat\Pi^{\mu\nu}(\bm{r}) =  \frac12 \sum_{i\neq j} [ \ff_{ij}^\mu \delta(\bm{r}-\bm{r}_j) + \ff_{ji}^\mu \delta(\bm{r}-\bm{r}_i) ]
\\ - \nabla_\nu \left[ \sum_i \frac{p_i^\mu {p}_i^\nu}{m} \delta(\bm{r} - \bm{r}_i ) \right] \; .
\end{multline}
Interchanging dummy indices in the first sum and plugging into (\ref{equ:IK-pdot}), one recovers (\ref{equ:p-bal}).
 
 We note that the IK stress is not the only choice for a tensor that satisfies~\eqref{equ:p-bal}, hence local stress is not uniquely defined, see for example~\citeParola.  However, the forces that we compute via stress gradients are unique.  (The difficulty with the stress itself arises because interparticle forces have finite range; a unique stress can be defined mesoscopically by averaging over scales much larger than the range of the force.) 
 
\subsection{Stress measurements in the biased ensemble}

 This section details the numerical procedure that was used to estimate the stress.

For numerical measurements, (\ref{equ:IK-stress}) is awkward because of the Dirac delta functions.
A similar issue arises when estimating the density from simulations: the solution in that case is to measure the density as a histogram. Specifically, one considers a region $\Omega$ of the system and defines
\beq
\hat{N}_\Omega = \int_\Omega \hat\rho(\bm{r}) d\bm{r} 
\eeq
which is equal to the number of particles in $\Omega$.  This is easily computed in simulations, and one may estimate the average local density at $\bm{r}$ as $\langle \hat{N}_\Omega \rangle / |\Omega|$ where $\Omega$ is a region centred at $\bm{r}$, and $|\Omega|$ is the volume of $\Omega$.  

A similar method can be applied to measure the stress (see for example~\citeSmith), we define
\begin{align}
\hat\Pi^{\mu\nu}_\Omega 
& = \frac{1}{|\Omega|} \int_\Omega \hat\Pi^{\mu\nu}(\bm{r}) d\bm{r} 
\nonumber \\ 
& =  \frac{1}{2|\Omega|} \sum_{i\neq j} ({r}_j^\nu - {r}_i^\nu ) H_\Omega(\bm{r}_i,\bm{r}_j)   F_{ij}^\mu - \frac{1}{|\Omega|} \sum_{i\in\Omega} \frac{p_i^\mu {p}_i^\nu}{m} 
\label{equ:IK-vol}
\end{align}
where $H_\Omega(\bm{r}_i,\bm{r}_j)=\int_\Omega h(\bm{r};\bm{r}_i,\bm{r}_j) d\bm{r}$ is the fraction of the line from $\bm{r}_i$ to $\bm{r}_j$ that passes through the volume $\Omega$, and the notation $i\in\Omega$ indicates that we sum over particles whose positions $\bm{r}_i$ are inside $\Omega$.  Then $ \hat\Pi^{\mu\nu}_\Omega  $ is the local stress, averaged over $\Omega$.  This is called the volume-averaged stress~\citeSmith.

In practice, we divide the system into square boxes of size $l_0/8$ and measure the stress in each box.  Results for the stress tensor are then obtained by averaging along the $y$ direction.   The definition (\ref{equ:IK-vol}) ensures that this procedure (measuring the stress in small boxes followed by averaging over the boxes) is equivalent to making the measurement on larger boxes from the outset.
\vfill

\end{appendix}


%

\bibliography{dev}
\end{document}